\documentclass[onecolumn,RNAAS]{aastex62}
\def\chandra{\textit{Chandra}}
\def\pg{\textit{PG~1116+215}}
\def\oviii{O~{\small VIII}}
\def\ovi{O~{\small VI}}

\received{January 1, 2018}
\revised{January 7, 2018}
\accepted{\today}
\submitjournal{RNAAS}

\shorttitle{New observation of \pg}
\shortauthors{Bonamente, M.}

\begin{document}

\title{New CHANDRA observations of \pg\ to investigate an extragalactic \oviii\ WHIM absorption line}

\correspondingauthor{Massimiliano Bonamente}
\email{bonamem@uah.edu}

\author{Massimiliano Bonamente}
\affiliation{University of Alabama in Huntsville,
Huntsville, AL 35899, USA}

\author{Jussi Ahoranta}
\affiliation{University of Helsinki, 
Helsinki, Finland}

\author{Jukka Nevalainen}
\affiliation{Tartu Observatory,
Tartu, Estonia}

\author{Patrick Holt}
\affiliation{University of Alabama in Huntsville,
Huntsville, AL 35899, USA}





\section{Introduction and Data Analysis} 
\label{sec:data}
\label{sec:intro}

In \cite{bonamente2016} we reported the  detection of \oviii\ absorption line from 
the $z=0.1765$ quasar \pg, associated with a  $z=0.0928$ \ovi\ system \citep{tilton2012},
and interpreted as warm--hot intergalactic medium (WHIM) along the sightline.
The report was based on a 90~ks \chandra\ HRC/LETG observation taken in 2002.
In 2018 we observed the source in the same \chandra\ configuration for an additional 280~ks.

\begin{figure}
\includegraphics[angle=0,width=5.0in]{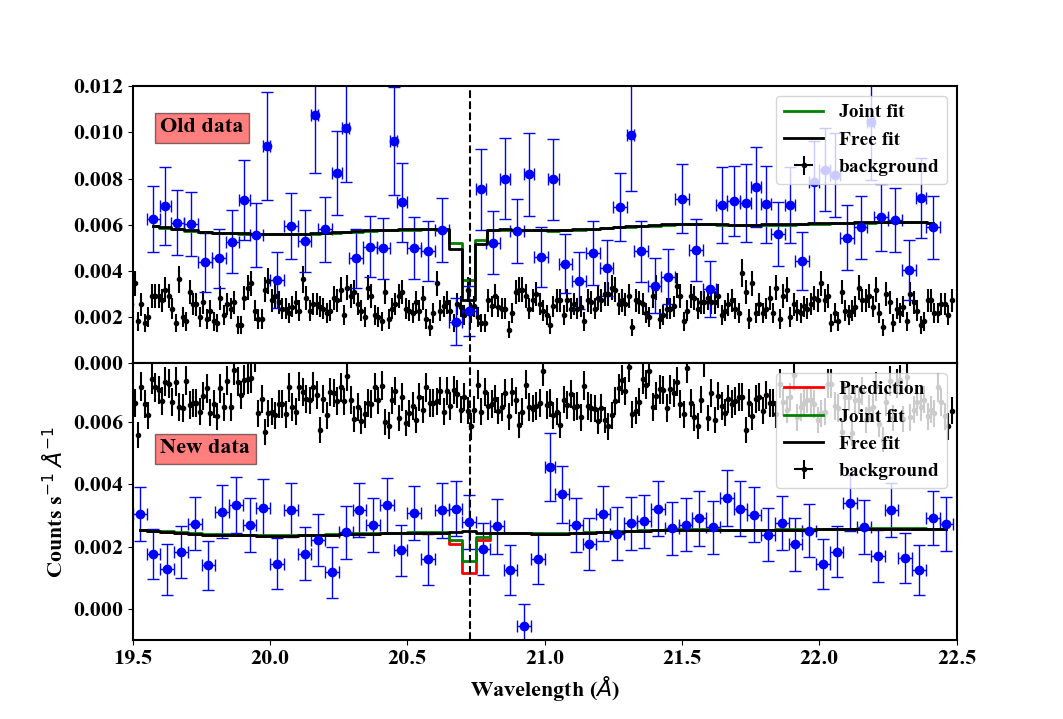}
\hspace{-1cm}
\includegraphics[width=2.6in]{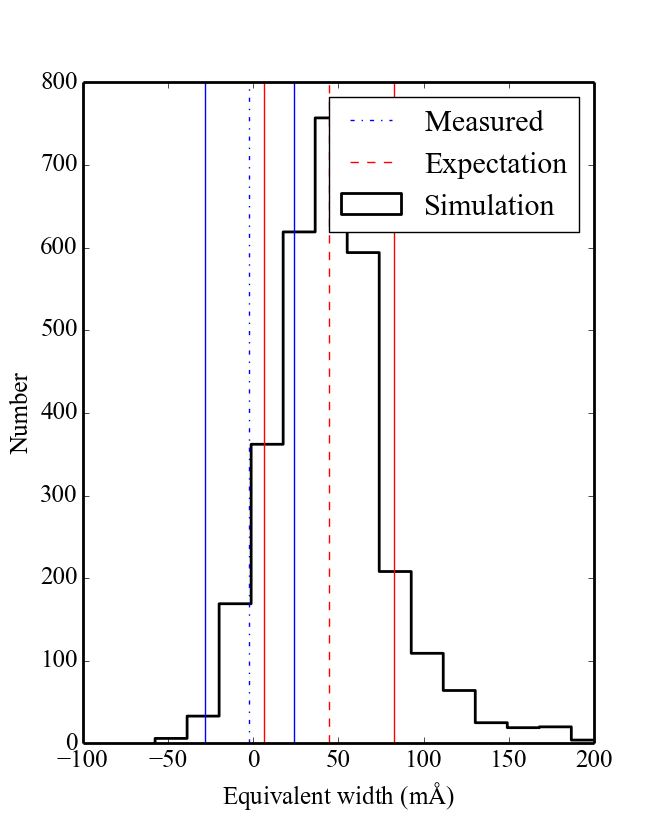}
\caption{(Left) \chandra\ HRC/LETG spectra from 2002  ("Old data") 
and 2018 ("New data"). The "Prediction" in the "New Data" panel is 
from the best--fit model of the "Old Data" \oviii\ line. In 2018, the flux was 
lower by a factor of $\sim$3 and the background higher by a factor of $\sim$3.
"Joint fit" is model from the simultaneous fit to old and new data, with linked \texttt{zgauss} models.
(Right) Distribution function of EW of the \oviii\ in the 2018 data. 
The "Expectation" includes uncertainties in the 2002 model of the line,
with
 input value
$W_{\lambda}=44.5\pm14.1$~m\AA.
The 2018 measurement is $W_{\lambda}=-2.1\pm26.0$~m\AA.}
\label{fig}
\end{figure}

The new and old data were reduced with the 
standard \chandra\ \texttt{CIAO 4.9} tools  \citep{fruscione2006}
to generate a single spectrum that combines the $\pm1$ order data.
The spectra were then analyzed in \texttt{XSPEC} \citep{arnaud1996}
 in the 19.5--22.5 \AA\ range, 
chosen to have a sufficient baseline to determine
the continuum, and narrow enough for an accurate modelling of the background with a simple power--law model
(Figure~\ref{fig}).
The two spectra were fit simultaneously, each with a 
power--law plus a \texttt{zgauss} model with fixed redshift $z=0.0928$, corresponding to
the \cite{tilton2012} \ovi\ detection. In the first fit, the 
\texttt{zgauss} normalizations are left free between the two models;
at the wavelength of the expected \oviii\ K$\alpha$ line ($\lambda\simeq20.65$~\AA),
the 2002 data show a significant absorption feature (\texttt{XSPEC} normalization $K=-17.0\pm5.6 \times 10^{-6}$)
while the 2018 data show no significant absorption ($K=0.4\pm5.0 \times 10^{-6}$), for a \emph{Cash} 
fit statistic  $C=507.1$ for 473 degrees of freedom.
In the second fit, the \texttt{zgauss} models were linked to ensure the same equivalent width for the two 
\oviii\ lines; the normalization
becomes $K=-12.0\pm5.3 \times 10^{-6}$ for the 2002 data (and $-6.0\times 10^{-6}$ for the
2018 data, lower than the 2002 data because of its lower flux) for $C=509.6$ with 474 d.o.f.~\footnote{In 
the original analysis of the 2002 data,
we reported $K=-21.1\pm5.8 \times 10^{-6}$,
differences may be due to changes in calibrations and in the choice of fitting bands.  We also reported a
free-redshift analysis of this line, in which the best--fit redshift was $z=0.0911\pm0.0004$.} 

In the new data the source intensity was \emph{fainter} by approximately a factor of two 
($F_{\lambda}=1.9\times 10^{-3}$ photons~s$^{-1}$~cm$^{-2}$~\AA\ in 2018, $F_{\lambda}=3.9\times 10^{-3}$ in 2002),
while the HRC/LETG first--order efficiency was also lower by $\sim 20$~\% compared to the 2002 data, leading
to an observed source flux $\sim$3 times lower in 2018.
Moreover, the background was also $\sim 3$ times \emph{higher} in the new 2018 data compared to the 2002 data, due to a known
anti--correlation with Solar cycle activity. In the 19.5--22.5 \AA\ band, the source accounts for 27.2\% of the total
counts for the 2018 data, and 69.8\% for the 2002 data. 


\section{Simulations to assess statistical consistency between 2002 and 2018 data}
\label{sec:simulations}
To investigate  whether the two datasets are consistent,
we simulated 3,000 
total and background spectra for the 2018 data, using the 
\oviii\ equivalent width predicted from
the 2002 data and its error. The redshift of the \oviii\ line was fixed at the nominal value.
We used an analytical model of the
background in the 19.5--22.5~\AA\ range; this model was used to simulate the background and the total spectrum using the \texttt{fakeit}
command in \texttt{XSPEC}, for each iteration of the simulation.
The equivalent width $W_{\lambda}$ was calculated according
to $W_{\lambda} I_{\lambda} = K$, negative values indicate excess emission
(Figure~\ref{fig}, right panel).


\section{Discussion and conclusions}

The simulations indicate that the \oviii\ line was expected to have an
equivalent width of $W_{\lambda}=44.7\pm38.2$~m\AA, based on the estimates from the 2002 observations.
The measurement was $W_{\lambda}=-2.1\pm26.0$~m\AA, i.e, the 2018
data did not provide a positive detection of the line.

We performed two quantitative tests of the consistency between the \oviii\ line in the 2002 and 2018 data.
First, the 2018 measurement of the EW corresponds to the 6.6 one--sided percentile 
of the expected distribution based on the 2002 data,
for a 6.6\% probability
to obtain such a small EW, or smaller.
Second,  the difference 
between the 2018 measurement and the expectation is $-46.8\pm46.3$~m\AA, i.e.,
the measurement falls just outside the 1--$\sigma$ interval. Assuming Gaussian distributions, there is a $\sim$32\% probability
for such value of the difference, or larger. 
Neither test can significantly exclude the hypothesis that the 2018 data are consistent with the 2002 data.
 We only used statistical (Poisson) errors in the
simulation of the spectra. The use of an additional systematic errors would bring the measurement
and expectation into an even better agreement.

We conclude that the 2018 observations of \pg\ are consistent with the 2002 observations, at the
wavelengths of the $z=0.0928$ \oviii\ K$\alpha$ line. The 
low flux of the source and higher background did not permit the confirmation of the
WHIM line made in \cite{bonamente2016} with greater confidence than
in the original data.

\bibliographystyle{aasjournal} 


\begin{thebibliography}{}
\expandafter\ifx\csname natexlab\endcsname\relax\def\natexlab#1{#1}\fi
\providecommand{\url}[1]{\href{#1}{#1}}
\providecommand{\dodoi}[1]{doi:~\href{http://doi.org/#1}{\nolinkurl{#1}}}
\providecommand{\doeprint}[1]{\href{http://ascl.net/#1}{\nolinkurl{http://ascl.net/#1}}}
\providecommand{\doarXiv}[1]{\href{https://arxiv.org/abs/#1}{\nolinkurl{https://arxiv.org/abs/#1}}}

\bibitem[{{Arnaud}(1996)}]{arnaud1996}
{Arnaud}, K.~A. 1996, in Astr. Data Analysis Software and Systems V, ed. G.~H.
  {Jacoby} \& J.~{Barnes}, Vol. 101, 17

\bibitem[{{Bonamente} {et~al.}(2016){Bonamente}, {Nevalainen}, {Tilton},
  {Liivam{\"a}gi}, {Tempel}, {Hein{\"a}m{\"a}ki}, \& {Fang}}]{bonamente2016}
{Bonamente}, M., {Nevalainen}, J., {Tilton}, E., {et~al.} 2016, \mnras, 457,
  4236, \dodoi{10.1093/mnras/stw285}

\bibitem[{{Fruscione} {et~al.}(2006){Fruscione}, {McDowell}, {Allen},
  {Brickhouse}, {Burke}, {Davis}, {Durham}, {Elvis}, {Galle}, {Harris},
  {Huenemoerder}, {Houck}, {Ishibashi}, {Karovska}, {Nicastro}, {Noble},
  {Nowak}, {Primini}, {Siemiginowska}, {Smith}, \& {Wise}}]{fruscione2006}
{Fruscione}, A., {McDowell}, J.~C., {Allen}, G.~E., {et~al.} 2006, in SPIE
  Conference Series, Vol. 6270, 62701V

\bibitem[{{Tilton} {et~al.}(2012){Tilton}, {Danforth}, {Shull}, \&
  {Ross}}]{tilton2012}
{Tilton}, E.~M., {Danforth}, C.~W., {Shull}, J.~M., \& {Ross}, T.~L. 2012,
  \apj, 759, 112, \dodoi{10.1088/0004-637X/759/2/112}

\end{thebibliography}

\end{document}